# How the Shale Revolution is Shaping the Future of the Oil and Gas Market

Binh T. Bui, PetroVietnam University
binhbt@pvu.edu.vn

## Abstract

The shale revolution, driven by advances in horizontal drilling, multi-stage hydraulic fracturing, and cyclic gas injection, has reshaped the oil and gas industry over the past two decades. In the United States, these technologies transformed ultra-low permeability shale formations into commercially viable resources, increasing crude oil production from 5 million bpd in 2008 to more than 12 million by 2019. Horizontal drilling became the standard after 2010, with nearly 200,000 horizontal wells completed by the end of 2023, accounting for more than 80% of all new wells in the US. This success is now being replicated in Argentina making the country a leading unconventional producer.

The common driver in both cases is the "mass manufacturing" approach, which industrializes oil production through standardized well designs, repeatable workflows, multi-well pad drilling, and continuous process optimization. Supported by a competitive oilfield service market and a skilled technical workforce, this model has reduced drilling times from months to weeks and cut well costs by half over the past decade. New technologies such as precision geosteering, advanced completion designs, real-time drilling analytics, and cyclic gas injection continue to improve efficiency and productivity.

Applying these methods and technologies to conventional reservoirs could significantly expand global production capacity and place sustained downward pressure on crude oil prices. This paper argues that in a period of slowing economic growth and potential financial deleveraging, abundant supply from both unconventional and conventional developments could extend a prolonged phase of relatively low oil prices. Such a scenario would have far-reaching implications for energy policy, investment strategies, and the competitive position of oil-producing nations.

## Introduction

The oil and gas industry continues to play a critical role in ensuring global energy security, despite the rapid growth of renewable energy. By the end of 2023, the three dominant fossil fuels still accounted for more than 77% of the world's total energy mix, with oil at 34%, coal at 28%, and natural gas at 23% (IEA, 2023). Global oil consumption currently averages about 100 million barrels per day, worth an estimated 6 to 8 billion USD each day, or roughly 2 to 3 trillion USD annually. This is approximately the entire GDP of Russia in 2023. Global natural gas consumption averages about 4 trillion cubic meters per year, worth an estimated 1.1 to 1.4 trillion USD annually, roughly equal to the entire GDP of Indonesia in 2023. The economic value of the oil and gas sector is reflected not only in its production volumes and revenues but also in its contribution to the global economy. Crude oil production alone represents roughly 2 to 3% of global GDP, a scale comparable to the entire global mining industry. Natural gas production adds another 1.5 to 2% of global GDP, depending on price fluctuations and annual output.

Despite the growth of renewable energy, oil and gas will not be replaceable in the near future. As of 2024, proven global crude oil reserves are estimated at roughly 1.7 trillion barrels, while natural gas reserves stand at about 6.9 quadrillion cubic feet; at current production rates, these resources could last another 50 to 60 years. However, extraction is becoming increasingly difficult as the most accessible fields are depleted and new large discoveries remain rare. One key indicator showing this trend is the Energy Return on Investment (EROI), the ratio of energy gained from a barrel of oil to the energy required to produce it. In the early days of the industry, average EROI values ranged from 50 to 100, meaning that the energy equivalent of one barrel could yield 50 to 100 barrels in return. By the 1970s, U.S. crude oil EROI had already fallen to around 30. For shale oil, it typically ranges between 5 and 10, meaning that producing one barrel requires the energy equivalent of 0.1 to 0.2 barrels, while Canadian oil sands are even lower, with an EROI of just 3 to 5 (Hall et al. 2009). Financial costs follow a similar trend: conventional fields once produced oil at $5 to $15 per barrel, whereas unconventional sources such as shale now require $40 to $70 per barrel. Stricter environmental regulations, covering wastewater treatment, emissions control, and site restoration, have further increased overall investment requirements. Consequently, the viability of oil and gas production increasingly depends on three pillars: technology, which determines recovery efficiency and cost competitiveness; policy, which shapes investment conditions and regulatory frameworks; and management, which influences operational execution, risk control, and long-term sustainability.

The United States is a prime example of how rapidly a nation can reshape its energy landscape. In less than two decades, it transitioned from being the world's largest net energy importer to a net exporter of crude oil and natural gas. This transformation was driven by the large-scale commercialization of shale resources, enabled by two pivotal technologies, namely horizontal drilling and multi-stage hydraulic fracturing. Together, these techniques unlocked vast ultra-low-permeability formations once considered uneconomic, propelling U.S. oil production to record highs while significantly reducing drilling and completion costs. Average well construction times fell from several weeks to one or two weeks, improving capital efficiency and accelerating project cycles.

Over the past two decades, the United States has seen a dramatic surge in oil and gas output. From 2008 to the present, crude oil production has climbed from about 5 million barrels per day to more than 12 million barrels per day (**Figure 1**), propelling the United States past Saudi Arabia and Russia to become the world's largest crude oil producer (EIA 2023). In 2023, average crude output reached 12.9 million barrels per day, maintaining its global lead over Saudi Arabia (~10.2 million b/d) and Russia (~9.4 million b/d). The United States is also the world's largest natural gas producer, with annual output of about 36 Tcf, roughly equal to the combined production of Russia (~22 Tcf/year) and Iran (~9 Tcf/year) (EIA 2024a).

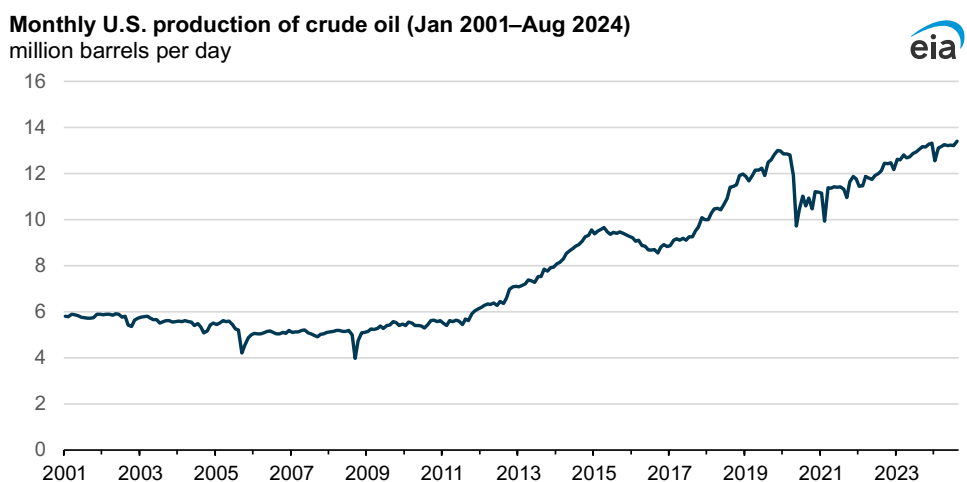

Figure 1. Annual U.S. crude oil production (EIA 2024a).

As of the end of 2023, the United States held an estimated 47.1 billion barrels of crude oil reserves and 374.3 Tcf of natural gas (EIA 2024b). In terms of oil reserves, the United States ranks 11th globally, significantly lower than Venezuela at approximately 303 billion barrels, Saudi Arabia at approximately 267 billion barrels, and Russia at approximately 108 billion barrels. For natural gas, United States reserves are approximately 22% of Russia's 1,688 Tcf and approximately 31% of Iran's 1,203 Tcf, which are the two largest reserve holders worldwide (BP 2023). Russia's crude oil reserves are more than double those of the United States, yet in 2023 its crude oil production of approximately 9.4 million barrels per day was about 73% of United States production. In natural gas, Russia's reserves are more than 4.5 times greater than those of the United States, although its annual output of approximately 22 Tcf is only about 61% of United States production.

A key distinction between the United States and the rest of the world lies in the composition of its reserves, with more than 60% of oil and nearly 75% of natural gas coming from unconventional sources, primarily shale oil and gas, which have very low per-well productivity and cannot be extracted using conventional technology. Major unconventional plays such as the Permian Basin (in Texas and New Mexico), the Bakken Formation (North Dakota), and the Marcellus-Utica Shale (Pennsylvania and Ohio) account for most of the country's unconventional oil and gas reserves. The Permian Basin is the largest shale oil-producing region in the United States, with estimated oil reserves exceeding 20 billion barrels and production of approximately 6 million barrels per day as of June 2024, representing about 40% of total U.S. oil output. The Bakken Formation is the second most important unconventional resource, with reserves of approximately 7.4 billion barrels. Although production there has declined somewhat in recent years, it still maintains an output of around 1.2 million barrels per day. The third key region includes the Eagle Ford and Marcellus-Utica shale plays. Eagle Ford currently produces nearly 1 million barrels per day, while the Marcellus-Utica is one of the largest natural gas fields in North America. Natural gas production in the Marcellus-Utica exceeds 35 Bcf/day, accounting for more than one-third of total U.S. natural gas output. The Marcellus Shale alone is considered the largest shale gas field in the United States, with reserves of more than 1,400 Tcf.

Notably, the U.S. oil and gas industry operates under a free-market system and is not controlled by state-owned enterprises as in other countries. As of 2023, more than 900 privately owned small and medium-sized companies were engaged in oil and gas production, particularly in shale oil and gas development. The United States also has a large and modern energy infrastructure, including more than 2.6 million km of oil and gas pipelines, over 120 refineries, and several major LNG export terminals such as Sabine Pass, Freeport, and Corpus Christi. This infrastructure not only ensures domestic energy security but also enables the United States to serve as a global energy supply hub. The development of infrastructure and the market-based model have been decisive factors in making the United States the world's leading oil and gas producer and exporter over the past two decades.

The U.S. shale experience highlights the decisive interplay between technological innovation, a transparent and competitive business environment, supportive regulatory frameworks, and a dynamic oilfield services sector. In the current energy era, these factors are as critical as, if not more critical than, resource endowment. The shale revolution has not only redefined U.S. energy markets but also offers valuable insights for oil and gas development globally. Accordingly, this paper aims to review the key enabling technologies behind the shale revolution and to discuss their potential implications for the future of the global oil and gas market.

## Unconventional resources

Unconventional resources are oil and gas deposits that differ from conventional reservoirs in terms of geology, technology, and economics, as they cannot be produced efficiently using traditional vertical drilling and natural reservoir pressure. Instead, they are trapped in low-permeability formations such as shale, tight sandstone, coal seams, or oil sands, requiring advanced extraction techniques like hydraulic fracturing, horizontal drilling, or thermal recovery. Economically, they are considered unconventional because their development typically involves higher costs and greater technological effort, although advances in technology and changes in market conditions can gradually shift them toward being more conventional. In broader energy discussions, the term may also extend to non-traditional sources such as oil shale or gas hydrates, highlighting the evolving nature of what is considered "unconventional".

In other words, unconventional reservoirs are formations where hydrocarbons cannot naturally flow into a vertical wellbore at commercially viable rates using the conventional method of lowering the bottomhole pressure (Bui 2016). Looking at the equation for flow rate toward a vertical well under steady-state conditions, it becomes clear that several factors can make production uneconomical, including high fluid viscosity, low reservoir permeability, thin formation thickness, depleted reservoir pressure, or a combination of these. In the literature, the most commonly cited examples are high-viscosity reservoirs (such as oil sands) and low-permeability formations (such as tight reservoirs).

A shale reservoir is a specific type of unconventional reservoir where oil or natural gas is generated and stored within fine-grained sedimentary rock called shale. Shale has extremely low permeability, meaning that hydrocarbons are trapped tightly within tiny pores and natural fractures. While shale is one type of tight formation, not all tight formations are shale. It is also important to note that not all U.S. unconventional production comes from shale formations. Therefore, the term "shale revolution" can be somewhat of a misnomer.

Despite significant technical challenges and relatively low profit margins, the global potential of unconventional oil and gas remains substantial. Unconventional oil reserves are estimated at around 419 billion barrels, while unconventional gas resources amount to roughly 7,576 Tcf. About 50 countries have been identified as having promising prospects and are already developing such resources. Major holders of unconventional gas reserves include China, Argentina, Algeria, the United States, and Canada, while leading unconventional oil reserves are found in the United States, Russia, China, Iran, and Argentina (**Figure 2**). These figures underscore the strategic importance of unconventional hydrocarbons, especially as reserves and production from conventional fields continue to decline.

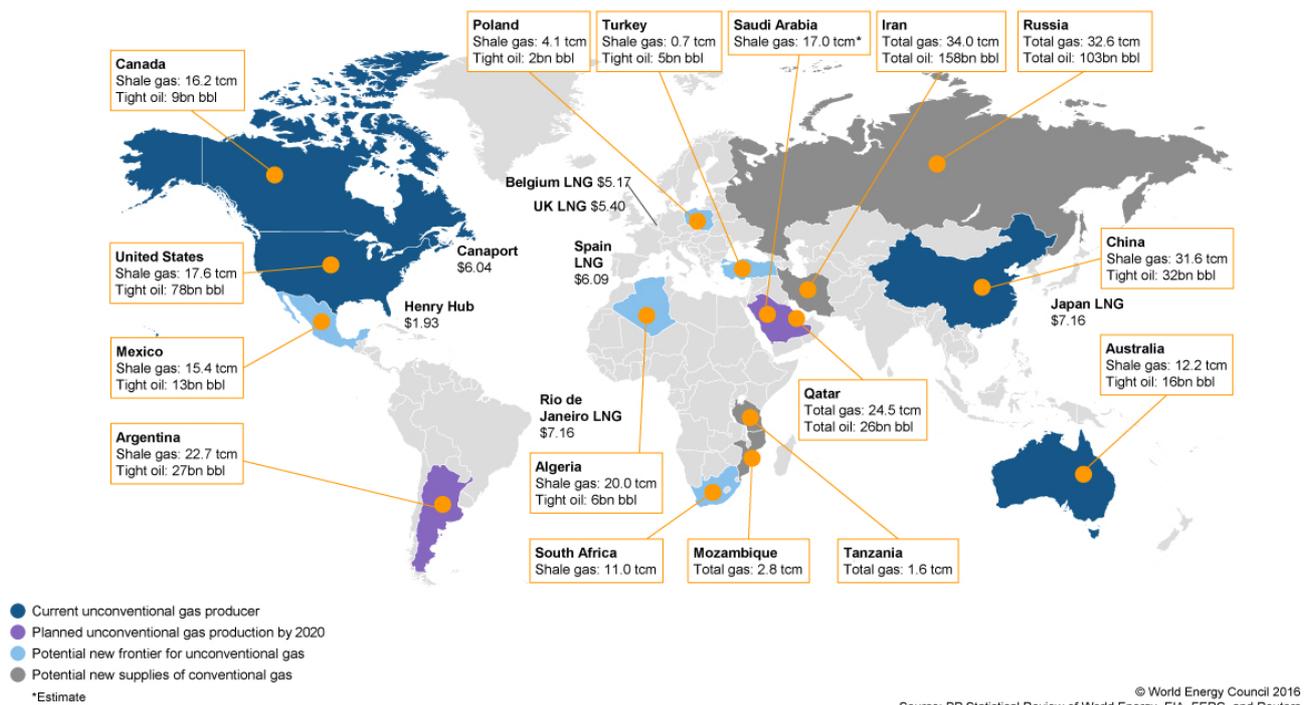

Figure 2. Global distribution of shale and tight oil reserves (World Energy Council, 2016).

The growth of the U.S. oil and gas industry over the past two decades has been driven by a convergence of factors, most notably policy, finance, and technology (Rignol and Bui 2020). First, a legal framework that allows private ownership of subsurface resources has created a favorable foundation, encouraging thousands of private companies to enter the sector. Second, following the early-2000s tech bubble, the U.S. government introduced a series of economic stimulus packages that channeled capital into shale production. Loose monetary policy helped sustain global economic growth, resulting in relatively steady oil and gas demand over the past three decades and providing a stable market for producers. At the same time, a well-developed and trusted financial market, with effective capital-raising tools such as equities, bonds, and venture capital, has played a critical role in funding high-risk exploration and production projects, particularly in shale oil and gas.

In addition, favorable tax policies, modern transport and pipeline infrastructure, a highly competitive market environment, and a culture of transparency have all contributed to a dynamic and efficient oil and gas ecosystem. However, the key factor in the long-term success of the U.S. petroleum industry has been systematic investment in research and development, particularly research that is rigorous, innovative, and practically applicable. Equally important are the breakthrough technological advancements, especially in horizontal drilling and multi-stage hydraulic fracturing, which have enabled the successful commercialization of shale oil and gas resources.

## Origin of shale revolution

After the oil crises of 1973 and 1979, the U.S. government recognized that reliance on imported oil, particularly from the Middle East, posed a serious threat to national energy security. In response, the United States launched a long-term strategy to build self-sufficiency in oil and gas production, with a strong focus on investing in advanced technologies to improve energy efficiency and boost oil production output. This strategy was implemented through the creation and funding of large-scale petroleum research programs, with particular emphasis on developing methods to extract tight oil and gas, resources whose reserves had been identified since the 1970s but were not economically recoverable with conventional technologies.

One of the pioneering initiatives was the Eastern Gas Shales Project (EGSP), launched in 1976 with funding from the U.S. Department of Energy (DOE). The program focused on studying shale formations in the Appalachian region, including the Marcellus and Utica Shales. Its activities encompassed exploratory drilling, core sampling, stratigraphic analysis, and testing reservoir stimulation methods such as hydraulic fracturing. The EGSP provided the scientific foundation and geological data that later underpinned the development of shale oil and gas extraction technologies. In parallel, the Gas Research Institute (GRI), a nonprofit organization established in 1976, became a key driver of applied research in the natural gas industry. Funded through a tax on domestic gas consumption, GRI invested hundreds of millions of dollars throughout the 1980s and 1990s. Its projects focused on developing low-cost multi-stage hydraulic fracturing techniques, efficient horizontal drilling methods, production monitoring systems, and, notably, supporting research in the Barnett Shale region of Texas, later recognized as the birthplace of the shale gas revolution. In addition, the National Energy Technology Laboratory (NETL), a research center under the DOE, has served as a central hub for developing, testing, and transferring advanced extraction technologies to the private sector. NETL's work has included not only research into multi-stage hydraulic fracturing and directional drilling, but also the creation of digital geological models to identify prospective zones within low-permeability formations.

Thanks to over two decades of sustained investment, by the early 2000s the United States had built a robust scientific and technical foundation capable of shifting from research to commercialization. Private companies such as Mitchell Energy successfully applied these advancements in practice, most notably by combining horizontal drilling with hydraulic fracturing to produce gas from the Barnett Shale. This milestone marked the beginning of the U.S. unconventional oil and gas revolution.

## Horizontal drilling

Since the late 19$^{th}$ century, numerous inventions have marked the evolution of directional drilling (**Figure 3**). The whipstock, first used in California in the 1890s, was the earliest tool for deflecting the drill bit and steering the wellbore. In 1929, H. John Eastman invented a magnetic surveying instrument, first applied at Huntington Beach, California, which made it possible to measure a well's inclination and azimuth, enabling precise directional drilling. By the 1940s, the development of the bent sub and downhole motor allowed for more flexible changes in drilling direction without relying entirely on drill string rotation. The combination of the whipstock, bent sub, and downhole motor formed the standardized directional drilling assembly that remains in widespread use today.

The introduction of measurement-while-drilling (MWD) systems in the 1970s improved drilling accuracy and speed by providing real-time data on well inclination and azimuth. In 1978, Teleco Oilfield Services, in collaboration with the U.S. Department of Energy, experimented with transmitting signals through drilling fluid, paving the way for continuous MWD technology in the early 1980s. MWD significantly reduced the time and cost of well surveying, enabled precise directional control, and laid the groundwork for horizontal drilling. In other words, measurement and directional control formed the foundation for the modern directional drilling. The major breakthrough in directional drilling came in the 1980s, when horizontal drilling was successfully applied in the Barnett Shale (Texas), enabling access to laterally oriented reservoirs. From the 1990s onward, rotary steerable systems (RSS) were developed, allowing continuous adjustments to the drilling direction without stopping drill string rotation, thereby providing precise control over the wellbore trajectory.

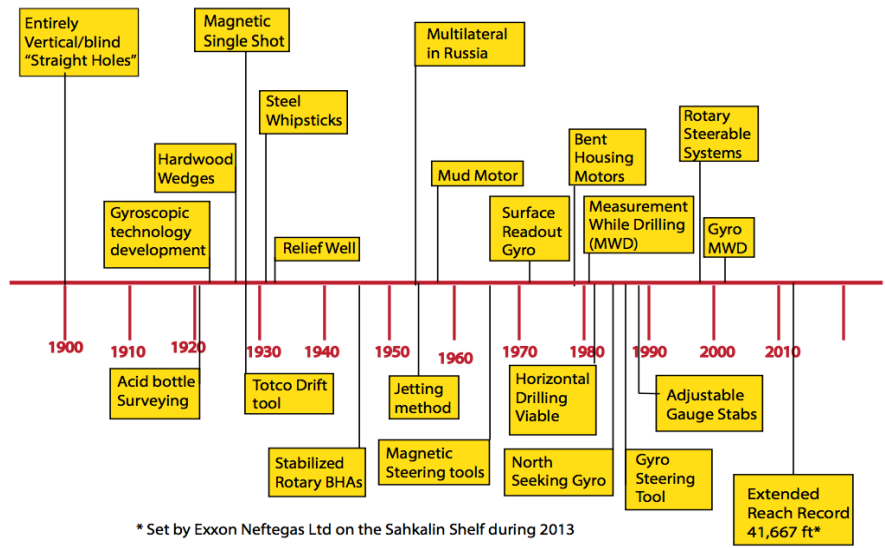

Figure 3. Important technology advancement in directional drilling (IADC 2015).

In addition to the well-known benefits of directional drilling, horizontal drilling extends the wellbore within the reservoir, increasing the flow area from the formation into the well by 20-30 times compared with vertical wells. Expanding this flow area not only boosts recovery but also enhances the effectiveness of multi-stage hydraulic fracturing by creating more extensive flow channels, significantly increasing production. As a result, the volume of oil or gas recovered from a single well can be many times higher while the cost increase remains relatively moderate. Although longer horizontal sections raise drilling costs, the cost per unit of lateral length decreases substantially (**Figure 4**) as the horizontal section lengthens. Consequently, the current trend is to drill wells with horizontal sections typically ranging from 3 to 5 km (**Figure 5**). This is the determining factor enabling unconventional oil and gas companies to reduce the costs.

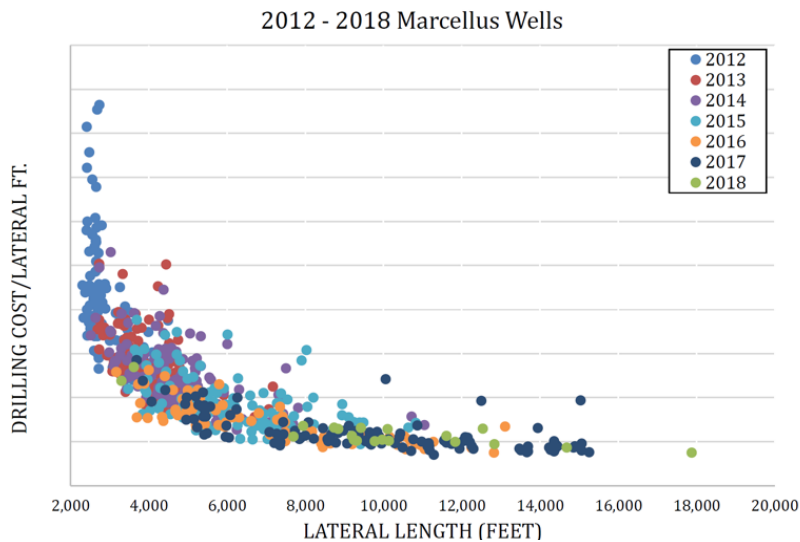

Figure 4. Variation of drilling cost per foot of the lateral with lateral length (Doak et al., 2018).

The United States has been a global pioneer and leader in horizontal drilling. Since 2010, horizontal drilling has become the standard method for shale oil and gas production, particularly in regions such as the Permian Basin, Marcellus, and Haynesville. Today, most new wells in the United States employ horizontal drilling combined with multi-stage hydraulic fracturing. By the end of 2023, nearly 200,000 horizontal wells had been drilled in the country. Although this is still a small fraction of the more than 4 million wells drilled across the United States to date (**Figure 6**), EIA data indicate that over 80% of all new wells in 2023 were horizontal (EIA 2024b). Horizontal drilling has thus become the dominant trend and a requirement for shale oil and gas development, a pattern that is likely to shape new well designs worldwide in the future.

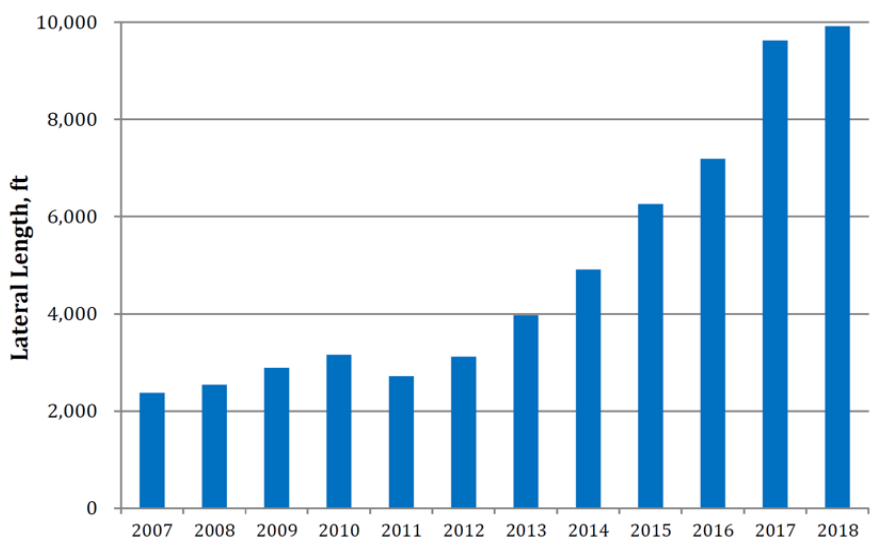

Figure 5. The evolution of lateral length in the Marcellus (Doak et al., 2018).

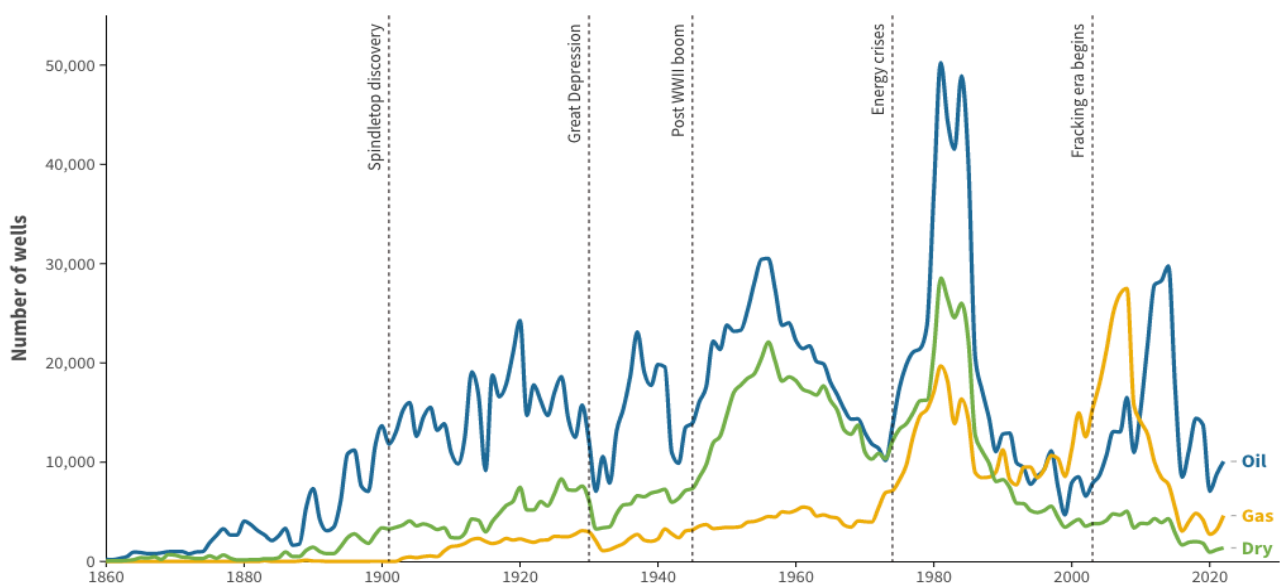

Figure 6. Number of new wells drilled in the U.S (IGS, 2025).

## Hydraulic Fracturing

Hydraulic fracturing involves using hydraulic energy to create fractures around a wellbore, increasing permeability so fluids can flow more easily to the well. The process begins by pumping a high-pressure fluid into the well, typically composed of water, proppant (usually sand), and a small amount of chemicals. The high pressure fractures the rock formation, and the proppant particles hold these fractures open once the pressure is released, allowing fluids to flow freely to the well. The fracturing fluid is primarily water (about 90-95%), combined with sand (5-9%) and a small proportion of chemicals (0.5-2%) that serve functions such as reducing friction, killing bacteria, and protecting equipment.

The idea of creating fractures in rock formations around a well to enhance production dates back a long time, but it only gained global attention in the early 2000s with the U.S. shale revolution. The history of fracturing began in 1862 when Edward Roberts recognized the potential of using explosives in oil extraction, applying the method to a well in 1865. This technique increased oil production from several wells by 1,200% in just one week. Explosive-induced wellbore fracturing became widespread throughout the late 19th and early 20th centuries. Today, this approach is still used and classified as explosive fracturing, though its application is limited due to safety regulations.

The modern form of hydraulic fracturing in use today began in the 1940s with research and testing by Floyd Farris at Stanolind Oil (now BP). In 1947, the first hydraulic fracturing experiment was conducted in southwestern Kansas, United States. Although it was not successful, it marked the beginning of modern hydraulic fracturing. By 1949, Halliburton successfully completed two commercial tests, paving the way for the commercialization of hydraulic fracturing in the 1960s and 1970s. Initially, water was used as the fracturing fluid, with sand as the proppant. By the 1970s, polymers were added to increase fluid viscosity and enable use in higher-temperature formations. During the oil price crisis of the 1970s, the U.S. government funded extensive hydraulic fracturing research, focusing on tight formations. Additives were introduced to adjust viscosity, reduce friction, kill bacteria, and prevent equipment corrosion, significantly improving the ability to carry proppants deep into fractures. Meanwhile, proppants such as natural sand, lightweight ceramics, and even aluminum were developed to keep fractures open after fracturing was completed.

During the 1980s and 1990s, advances in well measurement and monitoring equipment allowed engineers to better understand fracture propagation mechanisms. Pressure, temperature, and flow sensors, along with geological modeling, enabled real-time adjustments to fracturing pressure, improving both safety and efficiency. This period also saw the introduction of numerical simulation models to design fracturing programs, fluids, and proppants tailored to specific formations, using commercial software. These tools allowed engineers to simulate fracture geometry, pressure, pumping rates, and proppant effectiveness before field operations. In the late 1980s, slickwater, a low-viscosity fracturing fluid with friction-reducing additives, was first tested. Although adoption was limited at the time, slickwater later became a cornerstone of modern shale fracturing. By the late 1990s, multi-stage fracturing techniques were being developed, particularly for horizontal wells in shale formations. A notable example occurred in 1998, when Mitchell Energy successfully conducted a multi-stage hydraulic fracturing test in a horizontal well at the Barnett Shale, Texas, marking the beginning of the U.S. shale revolution..

In the past decade, multi-stage hydraulic fracturing technology has advanced significantly, with a focus on technical efficiency, automation, and environmental impact reduction. Drilling practices have shifted toward long horizontal wells with 30 to 60 stages, creating networks of fractures along the wellbore. Wells are often drilled in parallel from a single well pad (**Figure 7**), typically spaced over 650 ft apart, forming a network that enhances economic efficiency. Proppants have also evolved, with ultralightweight proppants, and resin-coated proppants gradually replacing conventional sands to maintain conductivity and fracture integrity under high in-situ stress. Meanwhile, fracturing fluids have been significantly improved. Next-generation slickwater fluids maintain high proppant transport efficiency while minimizing the use of harmful and expensive chemicals. Advanced monitoring technologies have also been widely adopted to optimize production. Microseismic monitoring is commonly used to determine fracture orientation and extent, while fiber-optic DAS/DTS systems provide real-time acoustic and temperature data within the well. In addition, digital monitoring platforms and real-time data analytics using artificial intelligence are increasingly applied, allowing fracturing operations to be adjusted flexibly and more efficiently.

Since its commercialization in the late 1990s, multi-stage hydraulic fracturing technology has been applied extensively across the United States. According to IPAA (2018), it is estimated that more than 2 million wells, both conventional and shale, have undergone fracturing to date. Most shale wells are hydraulically fractured, with an average of 10,000 to 20,000 new wells drilled and fractured each year, primarily in the Permian Basin, Eagle Ford, Bakken, and Marcellus Shale. IPAA (2018) reports that hydraulic fracturing has enabled the production of over 7 billion barrels of oil and 600 Tcf of natural gas. Additionally, a Harvard University study estimates that shale oil and gas development created approximately 2.7 million jobs in the United States during the first decade of the shale revolution. A 2013 study by the U.S. Chamber of Commerce projected that hydraulic fracturing could generate 3.5 million jobs nationwide by 2035.

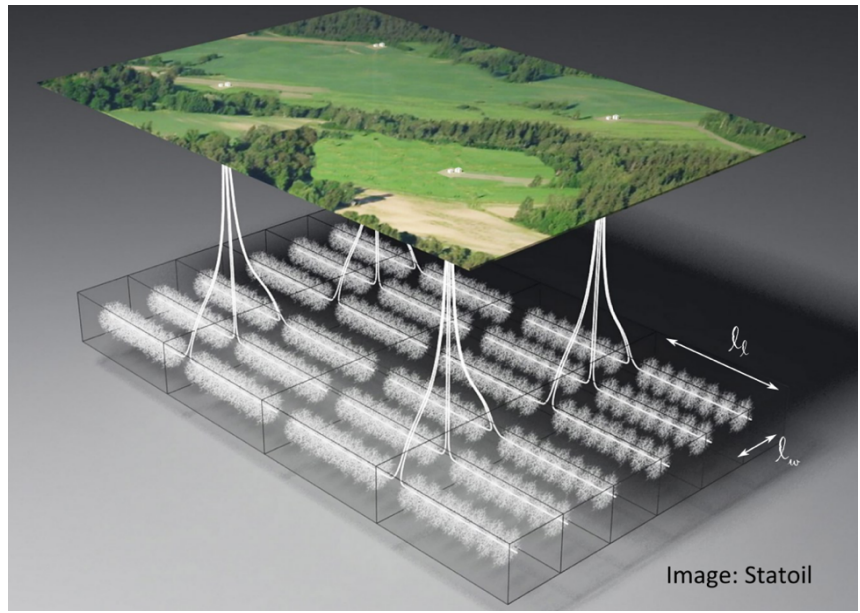

Figure 7. Sơ đồ bố trí giếng trong khai thác dầu khí đá phiến (Statoil.com).

For a successful hydraulic fracturing operation, each well typically consumes between 10,000 and 40,000 m³ of water and 1,500 to 3,000 tons of proppant. On average, U.S. fracturing operations use roughly 400 million m³ of water annually, equivalent to the yearly water consumption of Hanoi, the capital city of Vietnam. In terms of cost, hydraulic fracturing alone amounts to approximately $4-6 million per well, accounting for 40-60% of total well expenses. The cost structure varies depending on the well and the location. For example, at the Eagle Ford field (**Figure 8**), the largest cost component is completion equipment and materials, which represent about 40-50% of total costs. This is followed by well treatment services at 20-25%, proppant (frac sand) at 10-15%, and other expenses such as completion fluids and coil tubing.

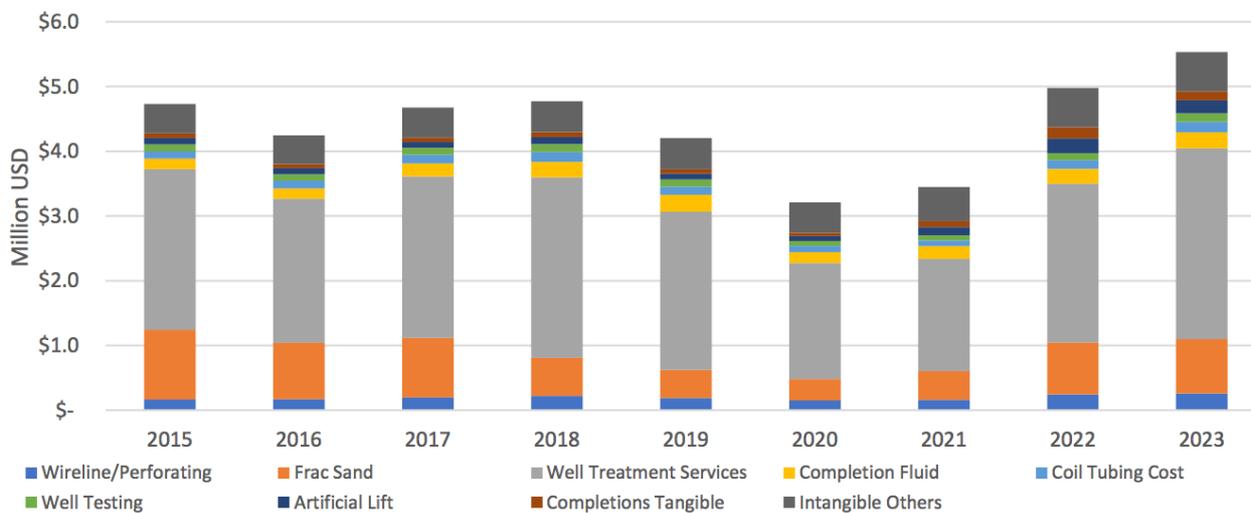

Figure 8. Hydraulic fracturing cost in the central Eagle Ford (Mubarak et al., 2019).

One of the key factors behind the success of hydraulic fracturing in the United States is not just the technology itself, but the organizational model and specialized service market that supports it. Unlike in many other countries, where fracturing technology is often confined to state-owned companies or large multinational corporations, the United States has dozens, if not hundreds, of private firms capable of independently supplying equipment, chemicals, personnel, and managing fracturing operations. A transparent, competitive market helps lower costs, accelerate deployment, drive continuous technological innovation, and easily scale operations. At the same time, the workforce structure is streamlined, consisting mainly of technical operators, with only a small proportion of engineers and supervisors, thanks to the standardization and automation of hydraulic fracturing processes.

## Enhanced oil recovery

Production from shale oil and gas wells declines rapidly, often dropping by 80-90% within the first one to two years because of the reservoir's physical characteristics. Even with modern technologies, recovery factors remain low, about 3-7% for oil and 20-30% for gas, compared with average recovery rates of roughly 30% for conventional oil and 60-80% for conventional gas (Bui and Tutuncu, 2017). This steep decline results in lower profit margins and shorter production lifespans, forcing operators to optimize costs. Given such low recovery factors, improving recovery through enhanced techniques is essential for the economic sustainability of shale development. Even a modest 1% increase in recovery can generate millions of dollars in additional revenue per well. According to the U.S. Energy Information Administration (EIA 2013), technically recoverable resources from shale total approximately 345 billion barrels. Using an estimated recovery factor of 7% for oil shale, a 1% increase in the global oil recovery factor could potentially unlock an additional 50 billion barrels, equivalent to 1.5 years of global oil production.

Recovery methods in shale reservoirs therefore extend beyond typical production practices, relying on innovative approaches that integrate advanced technologies and tailored strategies to improve recovery efficiency. Interestingly, many of these techniques, initially developed for unconventional resources, are now being adapted for conventional reservoirs. This suggests that shale enhanced oil recovery methods could provide valuable insights and technologies applicable to a broader range of oil and gas developments.

To enhance recovery efficiency, the cyclic gas injection process, commonly referred to as huff and puff has been extensively studied and applied over the past decade, particularly in major shale plays such as the Eagle Ford, Bakken, and Permian. The method is implemented in three stages: (1) injection (huff) - high-pressure gas is injected into the producing well over a period ranging from several days to weeks, (2) soaking - the well is shut in to allow the injected gas to diffuse into the reservoir matrix and mix with the oil, and (3) production (puff) - the well is brought back onstream, during which the pressure decline facilitates the recovery of the oil-gas mixture at the surface (**Figure 9**).

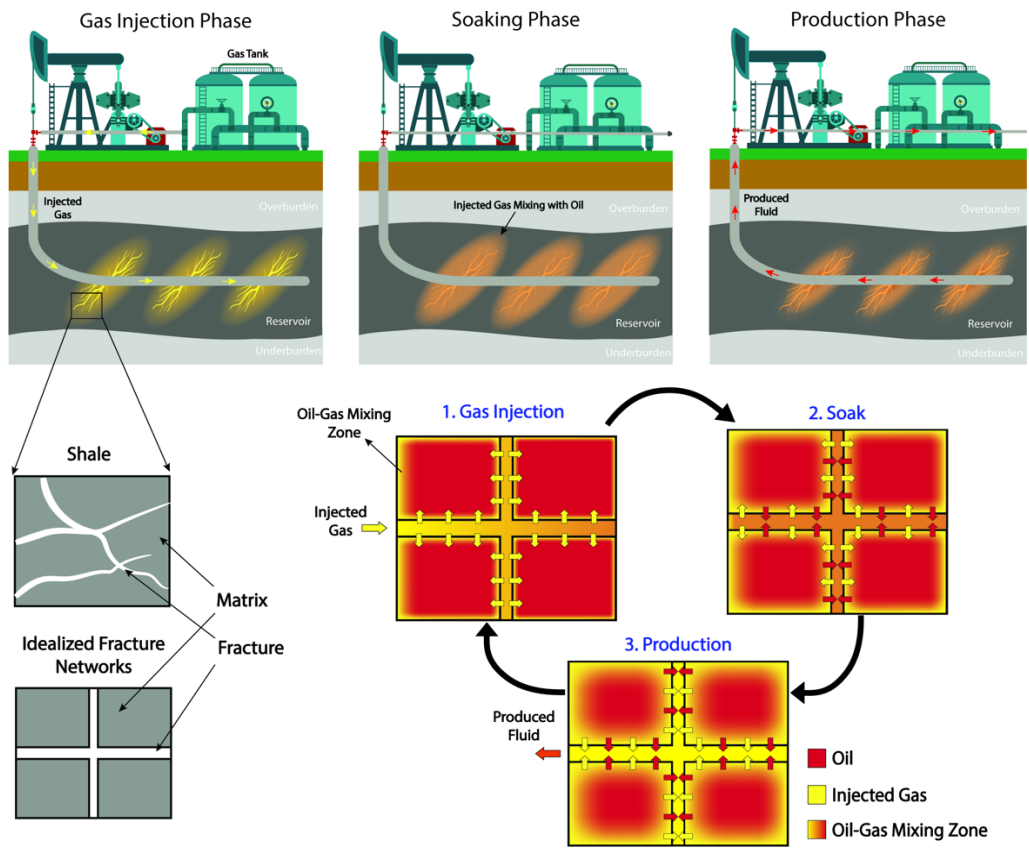

Figure 9. Ilusstration of huff-and-puff technique (Meybodi et al., 2024).

The most commonly utilized injection gases include $CO_2$, natural gas rich in light hydrocarbons, and liquefied petroleum gas. Among these, $CO_2$ is considered the most effective due to its ability to swell crude oil, reduce interfacial tension, and decrease viscosity, thereby enhancing oil mobility within tight reservoir rock structures. This technique can increase oil production by 30-70% compared to natural decline, with recovery factors reaching 15-20% of the original oil in place (OOIP) in certain cases. Additionally, it can extend well life by 3-5 years and improve economic returns by 20-50%. In terms of cost, each $CO_2$ injection cycle typically requires an investment of USD 0.5-1.5 million per well, depending on reservoir depth.

The huff-and-puff technique offers a significant advantage over conventional enhanced oil recovery methods in that it can be implemented in a single well without the need to drill additional injection wells, thereby substantially reducing initial capital investment. Moreover, this method enables selective application to wells with high potential, rather than requiring field-wide deployment as in traditional enhanced oil recovery strategies. Implementation costs are typically 30-50% lower compared to continuous interwell injection.

This reflects a broader trend in enhanced oil recovery development, shifting from field-scale applications toward well-centric approaches. The shift is largely driven by insights from shale oil development in the United States, where huff-and-puff operations have demonstrated strong economic efficiency when applied directly to individual wells. With the advancement of reservoir modeling technologies, intelligent sensing systems, and the integration of artificial intelligence for real-time data analytics, sing-well enhanced oil recovery is not only technically feasible but also represents a promising direction for revitalizing mature and depleted fields in Vietnam and worldwide.

One of the distinguishing features of enhanced oil recovery in shale reservoirs is the use of a single-well approach. Unlike conventional reservoirs, which typically employ multiple wells with separate injectors and producers, shale oil recovery often relies on individual wells that function as both injector and producer. This method is particularly well-suited to the low permeability formations, where traditional multi-well systems may be less effective. Single-well enhanced oil recovery techniques, such as the huff-and-puff method, enable targeted interventions that maximize recovery from individual wells without requiring extensive infrastructure or multi-well coordination. Beyond shale applications, this innovation has implications for conventional enhanced oil recovery. In mature or marginal conventional fields, single-well methods could offer a cost-effective alternative where drilling additional wells is not feasible. They may also prove beneficial in tight or poorly connected zones within conventional reservoirs, where fluid communication is limited. Additionally, single-well techniques can serve as simplified pilots for early-stage enhanced oil recovery evaluation, reducing upfront infrastructure investment. Advances in cycle design, injection fluids, and reservoir monitoring from shale-based enhanced oil recovery may further inform and enhance conventional enhanced oil recovery practices, particularly in optimizing recovery and improving real-time reservoir management.

### Implication for future oil and gas market

Since the early 1990s, advances in directional drilling have enabled the United States to deploy horizontal wells, extending the wellbore laterally within the reservoir and thereby increasing contact with oil- or gas-bearing rock. At the same time, multi-stage hydraulic fracturing technology has been applied to create hundreds of artificial fractures in the formation, boosting permeability and allowing hydrocarbons to flow toward the wellbore. Today, horizontal sections in shale wells typically range from 3 to 5 km in length, with hundreds of fractures along the well, enabling initial production rates of several hundred to over a thousand barrels per day.

Compared with conventional wells, shale wells are drilled much shorter time thanks to significant technological advancements. While a conventional well can take one to three months to complete (**Figure 10a**), a shale well typically requires only one to two weeks for drilling (**Figure 10b**). Total well completion, including hydraulic fracturing, usually takes two to three weeks, and often less than two weeks. This rapid pace is achieved through three main factors. First, modern drilling technology with precise directional control reduces the time required for the horizontal section. Second, multiple wells are drilled from a single pad, minimizing rig setup and relocation time. Third, efficiency improves over time as drilling teams gain experience and optimize procedures for each well. Additionally, shale wells often feature smaller wellbores with a single casing string, significantly reducing casing and cementing time and allowing continuous drilling operations.

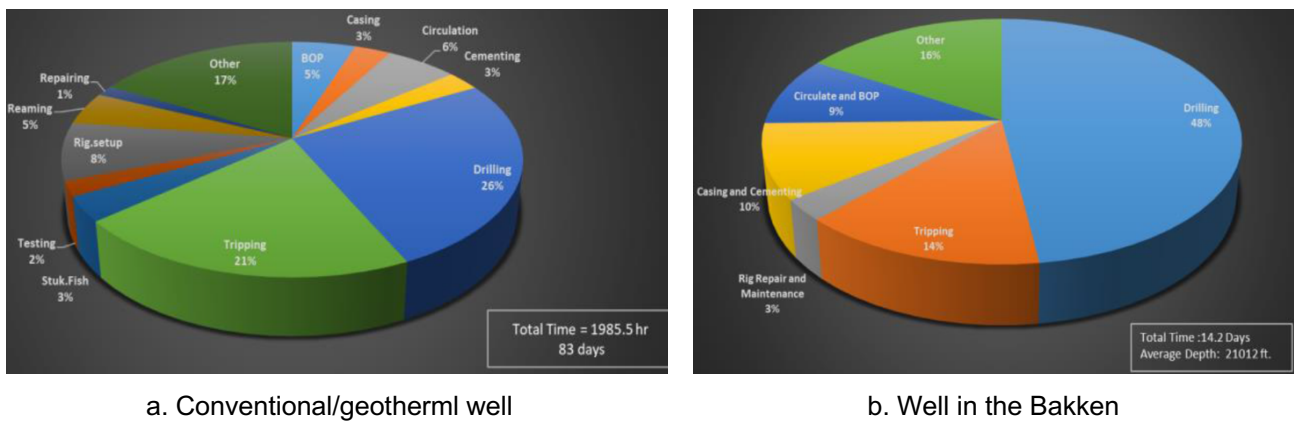

a. Conventional/geotherml well                b. Well in the Bakken

Figure 10. Drilling time distribution (AlMuhaideb and Noynaert, 2021).

Reducing drilling time is a key factor in significantly lowering costs for shale oil and gas operations. With rig rental and operating costs ranging from $30,000 to $100,000 per day, shorter drilling durations can save millions of dollars per well. The cost per foot drilled has also fallen sharply due to the use of smaller wellbores (**Figure 11**), dropping to just $0.20 per foot and bringing the drilling cost of a 7 km well length to approximately $5 million. According to EIA reports, the cost to drill and complete a shale well in the Permian Basin decreased from around $10 million in 2012 to roughly $6 million in 2023. Similarly, at the Eagle Ford (Texas), the total cost to drill and complete a 10,000-ft vertical well with a 9,600-ft horizontal section (total length about 6 km) is under $10 million, as summarized by Mubarak et al. (2019) in **Figure 12**.

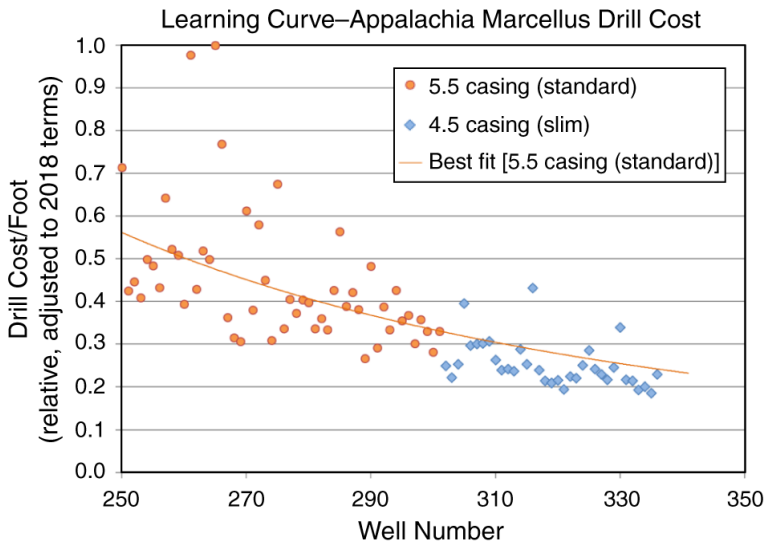

Figure 11. Drilling cost per foot at the Marcellus (Schumacker and Volgelsberg 2019).

Drilling cost reductions in shale have had profound implications for the industry, as they allow operators to remain profitable despite steep production decline rates and relatively low recovery factors. By lowering well costs through innovations such as pad drilling, walking rigs, advanced bit design, and real-time geosteering, shale producers have been able to reduce breakeven prices, improve capital efficiency, and drill more wells with the same investment. When these cost-saving practices are applied to conventional reservoirs, the impact can be equally significant. Lower drilling costs not only enhance the economics of conventional projects but also make smaller or previously marginal fields commercially viable. Techniques pioneered in shale, including faster drilling cycles, batch operations, and automation, can shorten development timelines and extend the productive life of mature fields by enabling cheaper infill drilling. As a result, transferring shale-driven efficiencies to conventional reservoirs not only boosts recovery economics but also unlocks additional resources, offering operators broader opportunities and greater sustainability across both unconventional and conventional assets.

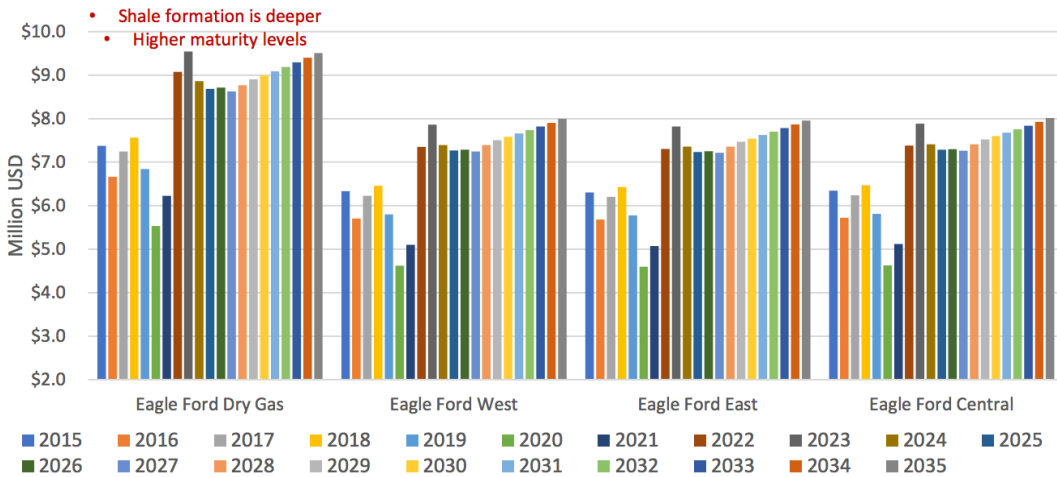

Figure 12. Chi phí khoan và hoàn thiện các giếng tại mỏ Eagle Ford (Mubarak et al. 2019).

One notable factor is the number of new wells drilled annually in the United States, which has exceeded 50,000 in some years (Figure 6). In recent years, the United States has drilled approximately 15,000-20,000 new oil and gas wells each year, accounting for over one-third of global new well activity (**Figure 13**) with the total cost of drilling and completing these wells is nearly equivalent to Vietnam's GDP in 2020. Although individual shale wells generally produce significantly less than conventional wells and experience faster decline rates, the large number of wells offsets overall production loss. Unlike traditional field development models, which focus on maximizing recovery across an entire reservoir, unconventional oil and gas production in the United States relies primarily on increasing well density and optimizing output at the individual well. This reflects a shift from geology-driven development, dominated by major and national companies, to an industrialized model in which many operators can participate. As a result, competition and innovation in technology and investment have been strongly encouraged.

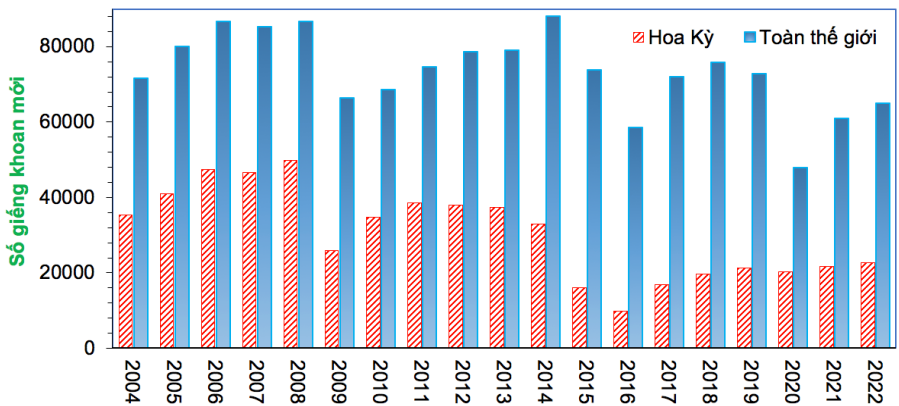

Figure 13. Number of new well drilled each year (Bui and Luong 2024).

The mass manufacturing approach is now being replicated in Argentina's Vaca Muerta formation, which already supplies about 400,000 bpd, nearly half of the country's oil output. By adopting U.S. drilling and completion practices, Argentina has accelerated and emerged as a leading unconventional producer. If the mass manufacturing approach were applied to conventional reservoirs globally, it could reshape the future of oil and gas production. Conventional reservoirs generally offer much higher recovery factors and longer production lifespans than shale wells, so scaling up drilling density and applying shale-style cost and efficiency improvements could dramatically increase overall recovery and accelerate production rates. This would not only make smaller and marginal conventional fields more economically viable but could also extend the life of mature reservoirs through cheaper infill drilling. Such a shift would represent a move away from traditional, geology-driven field development strategies toward a more industrialized, manufacturing-like model with broader participation, competition, and innovation. In effect, global oil and gas output could be expanded despite natural decline trends, as the system becomes less dependent on giant discoveries and more reliant on continuous drilling activity. Over time, pricing power for crude oil would shift from "leading exporting countries" toward technologically independent producers. This could lead to a rebalancing of global geopolitics. However, this approach would also bring challenges as it requires massive ongoing capital investment, robust service sector capacity, and highly ethical management. Still, the implications suggest that if shale-inspired drilling models were applied globally to conventional reservoirs, the industry could see both higher recovery and more resilient supply, fundamentally altering how oil and gas resources are developed in the future.

In addition to large-scale, industrialized development models, a key factor enabling the United States to maintain high drilling intensity is the existence of a broad and competitive oilfield services market. Today, hundreds of private companies in the United States possess the technical capability to drill complex horizontal wells. This creates a diversed service market that drives technological innovation, reduces costs, shortens project timelines, and avoids dependence on a few large firms. In contrast, in many other countries with centralized development models, only national oil companies or large multinational corporations have the financial, technical, and regulatory capacity to drilled horizontal wells. This concentration makes well deployment more complex, more expensive, and less flexible in responding to market fluctuations.

Another notable distinction of the U.S. model lies in the workforce structure for operations. In the United States, most well construction is carried out by trained technical workers rather than engineers, which reduces labor costs and enhances operational efficiency. In many other countries, however, drilling is still performed directly by engineers or highly specialized personnel, leading to higher labor costs and limiting the ability to rapidly scale up well numbers. For example, SM Energy, a U.S.-based independent shale producer, operates with a workforce of around 660 employees, about one-tenth the size of Vietsovpetro, the long-standing Vietnam-Russia joint venture, yet produces more oil. This stark contrast highlights the industrial efficiency of the U.S. shale model, where companies rely heavily on advanced technologies, outsourcing, and streamlined operations to maximize productivity. Vietsovpetro, by comparison, reflects a traditional, labor-intensive approach typical of national oil company structures, with broader responsibilities that include offshore logistics, marine services, and support functions that in the U.S. would often be outsourced. As a result, while both companies produce comparable volumes of oil, SM Energy achieves this with only a fraction of the workforce, underscoring the efficiency gap between industrialized shale development and conventional offshore operations. These differences not only illustrate disparities in technical expertise and economic conditions but also help explain why the U.S. unconventional oil and gas model can be deployed rapidly and at lower cost than in most other countries, despite geological and technical challenges.

The U.S. shale revolution has provided valuable lessons for conventional oil and gas production. First, horizontal drilling techniques have been optimized to achieve very long lateral sections, significantly increasing reservoir contact and boosting output. Second, multi-stage hydraulic fracturing creates hundreds of artificial fractures along the horizontal wellbore, greatly enhancing reservoir permeability and enabling hydrocarbons to flow to the well at commercial rates. Initially used primarily in ultra-low-permeability shale formations, multi-stage hydraulic fracturing is now also applied immediately after well completion in conventional reservoirs to increase production. This proactive fracturing of conventional wells has become a common practice to maximize recovery rates early in the well's life cycle, especially during periods of high oil prices. The implications of this trend for the future oil market are significant. By applying shale-derived technologies in conventional fields, operators can accelerate production, increase recovery factors, and improve project economics. Early and proactive fracturing of conventional wells not only captures more value at the start of production but also helps unlock bypassed hydrocarbons, extending the productive life of mature reservoirs. Widespread adoption of these practices could stabilize global oil supply, reduce reliance on new large-scale discoveries, and lower breakeven costs, making the industry more resilient to price volatility. In effect, the transfer of shale-inspired drilling and completion methods to conventional reservoirs signals a shift toward more industrialized, efficiency-driven oil development, with the potential to reshape competition and supply dynamics in the global oil market.

The United States has also developed a range of enhanced oil recovery technologies to address the steep production declines typical of unconventional reservoirs. Unlike conventional fields, where recovery is often boosted through large-scale injection networks, these efforts focus on optimizing recovery rates at individual wells. Techniques such as huff-and-puff gas injection, re-fracturing, and bottomhole pressure optimization are applied directly to single wells to extend productive lifespans and improve total recovered volumes. This well-by-well enhanced oil recovery model reflects a flexible approach that relies heavily on real-time production data and the unique geological characteristics of each well.

Moreover, the integration of digital technologies, including modeling, big data analytics, and artificial intelligence, has further enhanced operational optimization and decision-making. The implications of this innovation are far-reaching for the oil market. By tailoring recovery techniques to individual wells and leveraging digital tools, operators can sustain production longer, increase ultimate recovery, and reduce the cost per barrel produced. This flexibility allows shale producers to remain competitive even in volatile price environments, while also setting a precedent for conventional operators seeking to improve recovery efficiency without committing to expensive, field-wide projects. If widely adopted in conventional reservoirs, these shale-inspired approaches could lower decline rates, extend field life, and contribute to a more stable global oil supply. Ultimately, the combination of well-specific enhanced recovery and digital optimization represents a paradigm shift in reservoir management, one that could reshape both the economics and resilience of future oil production.

In the short term, an influx of new dollars into the financial system, combined with expansionary fiscal policies, can stimulate economic growth and push U.S. equity markets to new highs. Liquidity-driven momentum often boosts consumer spending, corporate earnings, and investment flows, creating strong demand across sectors, including energy. At the same time, supply chain redistribution and deglobalization trends are fueling large-scale investment in domestic manufacturing, infrastructure, and energy security. For oil and gas, this environment translates into stable short-term demand for hydrocarbons to support industrial activity and transport, while also reinforcing the role of U.S. shale as a flexible supplier capable of responding quickly to market signals.

Over the longer term, however, structural risks become more pronounced. Both the U.S. stock market and bond market are already trading at elevated levels, with signs of bubble dynamics amplified by years of ultra-low interest rates and liquidity injections. As stimulus fades and the debt services cost grows, inflationary pressures reemerge, or interest rates rise, financial markets could face sharp corrections. Deglobalization also introduces higher production costs and less efficient trade flows, which may limit global growth and suppress long-term energy demand. In this context, the oil and gas industry may experience periods of volatility: near-term opportunities for better prices could be followed by slower growth, tighter capital availability, and more cautious demand outlooks.

At the same time, it can be anticipated that technical advances from shale oil and gas production, particularly horizontal drilling, multi-stage hydraulic fracturing, and enhanced oil recovery methods, when widely applied to conventional fields, will significantly improve recovery factors and economic performance. In other words, the shale technologies when applied to conventional formations would greatly improve the productivity of the sector. These technologies not only optimize output from declining fields but also reduce per-barrel production costs, thereby contributing to an increase in global oil supply over the medium to long term. In addition, the technologies applied to shale can also be applied to other unconventional resources, such as high-viscosity oil, particularly Venezuelan oil, which could yield hundreds of billions of barrels to the market. In a context where energy demand growth is slowing due to looming financial market volatility and the ongoing shift toward "renewable energy", the broader adoption of such technologies is likely to place downward pressure on oil prices and increase price volatility in the future.

Ultimately, the convergence of financial market cycles, deglobalization, and technological transfer from shale underscores a paradox for oil and gas. In the short run, demand and prices may remain strong, buoyed by stimulus and industrial reshoring, but in the long run, higher efficiency, rising supply, and slower demand growth could create a more volatile, lower-priced environment. This dual outlook favors producers that combine cost efficiency and technological adaptability with financial resilience to weather rapid swings in the global energy market.

## Concluding remarks

The shale revolution has demonstrated how technological innovation and an industrialized approach to resource development can fundamentally alter the dynamics of the oil and gas industry. What began as a solution to unlock ultra-low-permeability reservoirs has evolved into a scalable model that is now being adapted for conventional fields worldwide. Horizontal drilling, multi-stage hydraulic fracturing, and sing-well enhanced oil recovery have redefined the technical and economic limits of hydrocarbon production, while digital technologies and process standardization continue to drive cost reductions and efficiency gains. These advances ensure that oil supply will remain abundant in the medium to long term.

Yet this abundance also brings new challenges. The broader adoption of shale-derived technologies in conventional reservoirs will likely place sustained downward pressure on oil prices and amplify volatility, particularly in an era marked by slowing global economic growth, financial market fragility, and the restructuring of supply chains under deglobalization. The result is a paradoxical outlook: while producers are better equipped than ever to extract hydrocarbons efficiently, the very success of these technologies may erode profitability by extending oversupply and constraining price recovery.

For policymakers, investors, and oil-producing nations, this evolving landscape demands strategic adaptation. Energy policy must account for both the resilience of supply and the risks of prolonged price weakness, while investment strategies should prioritize technological flexibility, cost efficiency, and financial resilience. Ultimately, the lessons of the shale revolution underscore that future competitiveness in the oil and gas sector will depend not simply on resource endowment, but on the ability to innovate, industrialize, and adapt to a more volatile and interconnected global energy system.